# 贸易政策不确定性对企业 ESG 表现的影响研究

**陈涵沁　陆 叶　黄华琴＊**

**摘要：**贸易政策不确定性已成为当今全球经济的重要特征之一，贸易政策不确定性在影响自由贸易的同时产生的微观经济效应如何仍有待商榷。为此，本文尝试探究贸易政策不确定性对企业 ESG 表现的影响及其作用机制，并利用 2010—2020 年 A 股上市公司数据展开检验。研究发现，贸易政策不确定性加剧能够显著而稳健地提升企业 ESG 表现。异质性分析发现，具备高新技术特性的企业更有能力在应对贸易政策不确定性时改善自身 ESG 表现，企业加强内部控制和任用具有环保背景的 CEO 也有助于其抓住贸易政策不确定性带来的机遇。从影响机制来看，贸易政策不确定性不仅能加剧企业所在的行业竞争程度，倒逼企业提升 ESG 表现以赢得市场份额，而且能够激励企业绿色技术创新产出，进而优化 ESG 表现。为此，应进一步完善 ESG 标准体系建设，建立 ESG 激励政策，提高贸易政策透明度和可预测性，推动企业绿色发展，助力国家整体可持续目标的实现。

＊ 陈涵沁，南京财经大学国际经贸学院；陆叶，中国外运长江有限公司苏州分公司；黄华琴，中国外运长江有限公司苏州分公司。通信作者及地址：陈涵沁，江苏省南京市南京财经大学国际经贸学院，210023

E-mail: 1220220031@stu.nufe.edu.cn

# 第一章　绪论

国际贸易仍存诸多不确定性。现今，在接连不断的冲击、日趋激烈的紧张局势和受损的多边贸易体系背景下，贸易政策不确定性持续上升，对经济全球化发展构成严重阻碍。自 2001 年加入 WTO 以来，我国对外贸易总额长期保持高速增长。与此同时，针对我国出口产品加征关税等贸易摩擦事件频发。截至 2019 年，我国已连续 24 年成为遭遇反倾销和反补贴调查最多的国家。随着贸易保护主义愈演愈烈，我国无数企业所面临的贸易政策不确定性（Trade Policy Uncertainty, TPU）增加，尤其是在特朗普政府上台之后，不断宣扬和实施激进的

反全球化和美国优先的贸易保护政策，使贸易环境不确定性对全球经济的影响引起广泛关注。

在全球经济持续低迷和贸易环境不稳定的局势下，ESG 是难得的坚定共识，也是不确定性的环境当中确定性的时代浪潮，对企业发展具有非常重要的意义。投资者和企业管理人员都应该继续坚持可持续发展的理念，持续对环境保护、社会贡献、公司治理绩效予以足够的重视。党的二十大报告突出高质量发展和绿色发展的重要性，为我国 ESG 发展奠定了理论与政策基石，也指引了我国未来 ESG 的方向。在创新、协调、绿色、开放、共享五大新发展理念驱动下，我国 ESG 呈现出强劲生长势头，其生态持续优化。近年来，为响应双碳目标、乡村振兴、共同富裕等国家战略，同时为配合监管要求，国内上市企业的 ESG 实践和信息披露水平明显提高。据 Wind 资料显示，A 股企业 ESG 管理实践平均得分从 2018 年的 4.09 分稳步提升至 2021 年的 4.6 分，2022 年受评级标准提高影响，平均分小幅回落 0.08 分，整体仍好于 2020 年水平。体现出 A 股在环境、社会及治理三个维度的管理实践能力逐年增强。在访谈中部分上市公司反映，采纳 ESG 实践不仅满足高质量发展的内在要求，还能显著提升公司管理效率。此现象说明，贸易政策不确定性虽会加剧企业外部环境波动性，但从发展视角来看，危机同样孕育着新机遇。

基于以上讨论，考虑到当前贸易环境愈加复杂，国内市场竞争愈加激烈，企业尤其是上市企业生存发展面临严峻的挑战，企业必须通过提升 ESG 表现不断增强自身的竞争优势。同时，提升 ESG 表现也是完成高质量发展宏伟蓝图的现实需要。为此，本文试图从企业层面验证贸易政策不确定性的微观经济效应，提出以下问题：第一，贸易政策不确定性能否显著推动企业 ESG 表现提升,其可能作用路径有哪些？第二，贸易政策不确定性是否通过加剧市场竞争进而影响企业 ESG 表现？企业绿色创新又是否在此过程中发挥中介作用？对于这些问题的研究将为我国企业应对政策不确定性冲击，支持引导企业 ESG 表现提升提供新的经验证据。

既有研究主要从战略决策、企业表现等微观角度，探讨了贸易政策不确定性对企业 ESG 表现的影响。就战略决策而言，宏观经济不确定性的外生变化对企业决策具有明确而有力的影响（Kumar S et al., 2023），在高不确定性时期，企

业的整体 ESG 绩效、环境绩效和治理绩效均有所提高（Vural-Yavaş Ç, 2021）。从投资者角度出发，尤其是长期、大规模、ESG 导向的投资者普遍认为，相对于撤资，积极主动参与风险管理是解决监管风险的更有途径（Krueger P et al., 2020）。就企业表现而言，当前关于贸易政策不确定性影响企业 ESG 表现的研究有限，主要聚焦于更高维度的经济政策不确定性如何影响企业表现。然而，众多研究中，关于政策不确定性如何影响企业 ESG 表现的结论并未达成一致。阻碍论一方认为，不确定性可能导致企业难以制定长期战略，对企业投资产生抑制作用（李凤羽，2015），进而降低其在 ESG 方面的投入。促进论则持有相反的观点，自愿发布社会责任报告的积极性显著增加，社会责任信息披露质量也显著提高（刘惠好和冯永佳，2020）。具体而言，虽然经济政策不确定性上升可能会加剧企业的系统性风险，但企业社会责任会在这一过程中发挥战略竞争工具的中介作用（阳镇，2021），进一步激励上市公司开展创新活动（顾夏铭，2018）。此举不仅是为减轻可能面临的风险冲击（王永海和郝晓敏，2022），也是在向利益相关者发送积极信号（Tiezhen Yuan et al., 2022）。

贸易政策不确定性这一抽象概念的测量问题、方法是理论与实证研究的核心与出发点。有部分文献选择直接测度贸易政策不确定性，例如，以中国加入 WTO 或其他特定贸易协定的冲击为视角，采用关税差额法衡量 TPU，将其定义为优惠关税可能回归至关税上限的风险（龚联梅和钱学锋，2018）。TPU 亦被量化为贸易协议中约束性关税承诺与实际关税间的差异，该差异反映出国际贸易环境的不确定性程度，进而影响企业及国家的贸易行动与策略（Osnago A et al., 2015; Carballo J et al., 2022）。也有研究通过贸易政策冲击事件作为准自然实验，例如永久正常贸易关系（PNTR）政策（毛其淋，2020）、中国加入 WTO（周定根等，2019；Zhihao Yang & Junjie Hong, 2021）、美国对华“301 调查”（罗宏等，2023）来探究贸易政策不确定性的变动效应。考虑到关税差额法无法覆盖其他非关税障碍效应，并且单一特定的准自然实验，如中国加入 WTO 事件也难以充分反映贸易争端加剧情况。于是，本文进一步重点关注整体刻画贸易政策不确定性及其经济影响的有关文献。当前主流做法是根据报纸报道频次并通过文本分析方法来构建政策不确定性指标，经过人工阅读在内的多种证据表明该指标能够反映政策相关的经济不确定性变化（Baker S R et al., 2016）。就国内而言，Huang 和 Luk（2020）

基于以上方法，采用文本分析计算在特定期间内，如北京青年报、广州日报等样本报纸中提及“贸易”、“政策”、“进出口”等关键词及其同义词的文章数量，并通过标准化处理获得 TPU 指标。相对而言，基于新闻文本的 TPU 指数能有效捕捉到国内政治及全球宏观波动引致的贸易政策不确定性，随着中国面临的贸易保护措施日益复杂化，文本提取法生成的 TPU 指数信息范围更广（郭平和胡君，2023），展现出其持续性、动态性及全面性特质，可观察更长时期 TPU 波动。

在已有研究基础上，本文边际贡献主要体现在两个方面：（1）在研究视角方面，本文站在贸易政策不确定性的视角，揭示特定来源不确定性对企业 ESG 表现的“倒逼”作用。通过这一研究视角，丰富对企业 ESG 表现影响因素的认识，也将 ESG 的影响因素探索扩展至更宏观视角，为理解 ESG 表现的多元影响因素提供了新的视角和思路。（2）本文于机制分析部分通过宏微观渠道探讨贸易政策不确定性对企业 ESG 表现的正向作用机制，并结合企业内控和领导层因素，提供丰富的异质性信息，能够深化对不确定性影响的理解，并为该领域提供多角度探索。这不仅为政策制定者和企业管理者提供了更多战略选择的思考空间，也为中国引导实体经济发展提供有价值启示。

本文接下来的安排如下：第二部分是理论分析，第三部分是研究设计，第四部分是实证结果分析，第五部分是异质性检验与机制分析，最后一部分为主要结论与政策启示。

# 第二章　理论分析

本章以战略性增长期权理论、动态能力理论和创新理论及相关研究为基础，明晰贸易政策不确定性对企业 ESG 表现的作用机制，进而为后文机制检验提供基础理论支撑。第一，本节从整体视角分析贸易政策不确定性对企业 ESG 表现的基础影响。第二，从行业竞争和绿色创新两个视角出发，厘清贸易政策不确定性影响企业 ESG 表现的内在机制。贸易政策不确定性影响企业 ESG 表现的具体路径如下方图 2.1 所示：

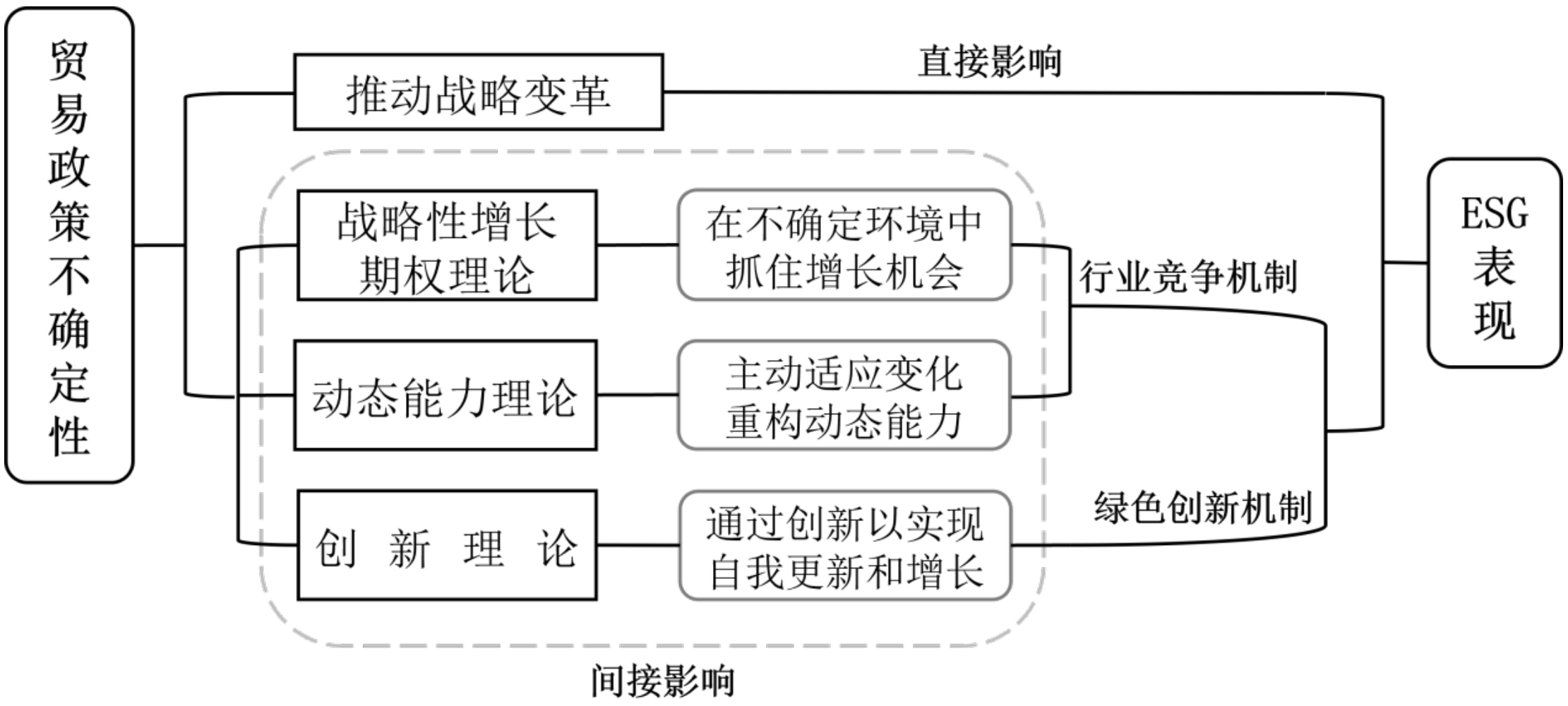

图 2.1 理论框架图

## 2.1 贸易政策不确定性与企业 ESG 表现

战略与环境的动态匹配过程要求企业在环境变化面前不断更新其战略资源配置，并增强对环境变动的适应性及内部协调能力，以在不断变化且复杂的环境中保持长久竞争优势。因此，战略变革的决策与执行效果会显著受到外部环境不确定性的影响（Eisenhardt K M & Martin J A, 2000）。

在资源受限且成本受控的前提下，理性企业致力于实现收益最大化，而企业是否选择在贸易政策不确定性时期提升其 ESG 表现，主要取决于其行为的成本效益分析。已有文献指出，政策不确定性与企业行为之间存在密切关联。贸易政策不确定性是企业无法预测的贸易政策波动，通常源于国际制度环境的变化，这种外部环境的制约迫使企业调整战略以适应新的发展要求。根据企业进化论和制度理论，贸易政策不确定性会推动企业战略变革，提升 ESG 表现是企业适应动态、复杂环境，并在此环境下保持竞争优势的重要手段。贸易政策不确定性主要通过以下两个方面直接影响企业 ESG 表现：一方面，基于企业进化论理论，贸易政策不确定性被视为外部环境中的一种压力，促使企业通过变异（创新战略和实践）、选择（市场对这些创新的测试）和保留（将成功策略固化为企业常规）的过程以不断适应外部环境不确定性。企业在面对贸易政策不确定性时，必须不断探索和实施新的战略变革以寻找最佳适应路径，进而在动态变化的贸易环境中生存和发展；另一方面，依据制度理论，企业在外部制约下，其组织结构和流程

趋向于获得意义并实现自身稳定（Lincoln J R, 1995），因此，企业通常会结合特定外部环境来制定和实施决策（Miles R E et al., 1978）。

在贸易政策不确定环境下，各方面的不稳定、不确定、复杂性和模糊性会给企业的生存、竞争和发展带来巨大冲击。面对复杂易变的贸易市场环境，很多企业会陷入焦虑不安，此时单一的企业变革手段和固有的企业变革管理模式难以使其更好地应对未来环境变化对组织发展提出的挑战，唯有进行战略变革、采取各种变革措施来应对这种变化才是最优解。

## 2.2 贸易政策不确定性、行业竞争与企业 ESG 表现

贸易政策不确定性带来的主要影响之一体现在国内行业竞争。随着贸易政策不确定性不断增加，国际市场经历剧烈波动，国内企业将会在出口市场面临上行压力。这种压力迫使企业重新定位，转而将市场重心放在国内，愈发加剧国内行业竞争环境。本文认为，可以从市场重心转移角度分析这一现象：第一，市场策略重新定位。贸易政策不确定性会引发国际贸易市场动态变化，导致部分企业为规避预期风险而选择退出市场，或使潜在新企业对进入市场持谨慎态度，通常涉及企业从风险较高的国际市场向相对稳定的国内市场转移。就出口市场而言，贸易壁垒高筑会显著减少企业在出口市场的进入行为，同时，显著增加已经存在于出口市场企业的退出行为，从而影响进出口贸易（Crowley M et al., 2018）。于是，贸易动态变化就会影响国内市场结构和竞争格局，进而导致国内剩余企业之间的竞争加剧。第二，供应链重组和调整。贸易政策不确定性的负面影响会通过生产网络蔓延到上下游企业（Yi Huang et al., 2023），导致供应链中断，促使企业寻求更加稳定的供应链解决方案，如转向或加强对国内供应链的依赖。随着企业对国内市场的依赖加深，国内行业的竞争压力随之增加。此外，供应链重组可能导致企业成本结构的变动，迫使企业在成本效率和供应链稳定性之间寻找平衡点，进一步激化行业竞争态势。

进一步，根据前文的理论基础，理解行业竞争的作用机制可从如下两个角度出发：一方面，从战略性增长期权理论角度，在行业竞争加剧的环境中，提升 ESG 表现可以被看作是一种特殊的增长期权投资，可以作为差异化策略，建立起竞争壁垒。同时，战略性增长期权理论强调长期价值创造，提升 ESG 表现正好将促进企业的持续增长和价值创造。在动态能力理论的框架下，企业面对不断

加剧的行业竞争及外部环境中的经济、社会、政治和技术维度的迅猛演进与多元化变革，须在其组织架构内部诸要素或行为主体之间培养出一种灵活而有效的应对及利用外部变动的能力。企业必须充分吸纳外部环境所提供的利好因素，同时转变那些与环境不再匹配的组织结构、运营模式及发展路径。通过这一过程，企业能够逐步形成动态的适应能力，这种能力是企业塑造自身独有核心竞争力的基石；另一方面，在动态能力理论框架下，企业面对日益加剧的行业竞争以及外界环境中经济、社会、政治和技术因素的快速发展及多元化变革，须在其组织架构内部诸要素或行为主体之间培养出一种灵活而有效的应对及利用外部变动的能力（Heckelman W, 2017），充分吸纳外部环境中的有利因素，同时改变自身与环境不再匹配的组织结构、运营模式及发展路径，促使形成动态适应能力，进而形成自身独特的核心竞争能力。综上，随着越来越多的企业将重点转向国内市场，国内行业竞争程度相应提高。在这种环境下，企业为了在竞争中生存和发展，就会寻求通过提高其 ESG 表现来彰显自身比较优势。

## 2.3 贸易政策不确定性、绿色创新与企业 ESG 表现

绿色创新属于企业平衡经济利益与资源环境的创新行为，能够满足当前我国对经济高质量发展的迫切需要。绿色创新区别于一般创新活动，主要是指绿色技术创新。绿色技术，作为减少环境污染、节约原材料和能源使用的技术、工艺或产品的集合，其与一般创新活动相比，需要更多的资金投入、更强的人才支撑和更高的技术水平。因此，绿色技术创新的特点是投入沉没大、过程不可逆及更显著的产出不确定，是一种长周期、高风险的创新行为。

在熊彼特的创新理论框架内，创新被视为推动经济发展和企业竞争力的关键力量。企业通过新产品、新技术、新市场开发、新供应来源的确立，以及新的组织管理方法的实施，引领市场和社会进步。同时，自然资源基础观中也包含类似观点，即绿色创新战略能够使企业拥有稀缺的、不可模仿的、不可替代的战略资源或能力，从而获得持续竞争优势（Hart S L, 1995）。

纵观众多关于政策不确定性影响微观经济活动的文献，在贸易形势不稳定性增加的情况下，企业通常会出于实现自我更新和增长的动机，加大绿色投入，加强绿色技术创新。从理论角度出发，从绿色创新视角探讨贸易政策不确定性对企业 ESG 表现的影响机制，可以从以下两阶段展开：

第一阶段，贸易政策不确定性会刺激企业绿色创新。贸易政策不确定性为企业提供一种特殊的外部经济风险，激励企业重新评估其经营策略和市场定位，从而寻找新的增长点和差异化竞争优势。在这种情况下，绿色创新成为企业响应这一不确定性和实现可持续发展的重要路径。具体而言，绿色创新能够帮助企业减少对不稳定贸易环境的依赖，通过提高资源效率、减少环境污染、降低能耗等方式，提升企业的适应能力和长期竞争力。

第二阶段，绿色创新不仅是企业适应贸易政策不确定性的战略选择，更是推动企业 ESG 表现提升的关键因素。一方面，根据知识基础观理论，创新能力决定企业表现，将企业特有的资源重新组合能为企业带来可观收益（Grant R M, 1996; Schumpeter J A et al., 2021）；另一方面，从实践角度出发，企业在创新过程中融入绿色理念，可以为其带来科技进步和节能减排双重效益，实现经济增长与生态环境保护共赢。较之末端强化的治理模式，绿色技术创新更有利于增强企业生态环境保护意识，能够帮助企业履行更多的社会责任，并推动企业管理效率提高、组织结构扁平化，从而提高公司治理能力，全面提升企业 ESG 表现（张萌和宋顺林，2023）。

# 第三章　研究设计

## 3.1 样本与数据

考虑到数据可获得性及现有相关研究，本研究选择 2010 年至 2020 年期间 A 股主板上市企业为研究样本，将我国 A 股上市公司的微观数据与贸易政策不确定性指数匹配，研究数据均来源于 CSMAR 数据库及公开可获得的网站 [1]。在处理上市公司数据库时，本文进行以下样本筛选与处理步骤：（1）剔除金融行业的企业，仅保留归属于实体经济领域的企业；（2）剔除股票交易中经过特殊处理（特指股票代码中含有 ST 和 ST*）以及主要财务指标缺失的企业，以保证本

[1] 贸易政策不确定性指标（TPU）来源网址为：https://economicpolicyuncertaintyinchina.weebly.com/。该网页展示我国大陆地区和特别行政区经济政策不确定性指数及其分项情况（包括财政政策不确定性、货币政策不确定性、贸易政策不确定性、汇率政策不确定性等指标）。

研究样本较高的数据完整性；（3）剔除样本期内重要指标缺失的企业；（4）为有效减轻极端值对数据分析的干扰，本文对所有连续变量进行 1%的双边缩尾处理。最终样本包含 1719 家企业，共 18909 个企业-年份观测值。

## 3.2 模型构建与变量选取

本文中被解释变量为企业 ESG 评级，属于从 1 至 9 分逐级递增的有序变量，并且在统计上被归类为离散变量。针对该类变量，本文选择运用有序 Logit 回归模型（Ordered Logistic Regression, Ologit）进行分析。同时，为进一步验证模型适用性，本文对样本数据进行豪斯曼检验，检验结果表明，应该选取固定效应模型来估计贸易政策不确定性对企业 ESG 评级的影响。首先，有序 Logit 回归模型表达式为：

$$P(y = k \mid X_n) = \frac{1}{1+e^{-(\alpha+\beta X_n)}} \tag{3.1}$$

其中，$X_n$ 表示第 n 个指标，y 代表企业 ESG 评级赋分高低（k = 1、2、3、4、5、6、7、8、9）。进而建立累积 Logit 模型：

$$\text{Logit}(P_k)=\ln[P(y\le k) / P(y\ge k+1)] = \alpha+\beta_1 L.TPU+X'\Gamma+\omega_t+\eta_i+\varepsilon_{it} \tag{3.2}$$

其中，$ESG_{ijt}$表示 i 行业的企业 j 在第 t 年的 ESG 表现；考虑到贸易政策影响的滞后性，同时为削弱潜在内生性问题的干扰，本文采用滞后一期的贸易政策不确定性指标（*L.TPU*）作为核心解释变量，表示我国第 t-1 年的贸易政策不确定性；X 包含一系列控制变量，包括企业年龄、资产负债率、总资产增长率、固定资产比率、股权集中度、总资产净利润率等；此外，模型加入年份固定效应（$\omega_t$）和行业固定效应（$\eta_i$），以排除行业层面不随时间变化的因素及时间趋势影响。具体而言，本文所选取变量如下：

（1）被解释变量

本文选取华证 ESG 评级数据作为分析企业 ESG 表现的数据源。华证 ESG 评级优势在于其紧贴中国市场特性，覆盖面广泛，更新及时，因此本文利用华证

指数针对沪深 300 指数成份股进行的 ESG 以及各相关维度的评级，来作为评估企业 ESG 表现的依据。在这个框架下，沪深 300 指数的成份股按 ESG 表现被分为九个级别，级别设置从高到低为 AAA、AA、A、BBB、BB、B、CCC、CC 到 C。为便于分析，本文将这九个级别按顺序赋值为 1 到 9 分，即从 C 级的 1 分到 AAA 级的 9 分。

（2）核心解释变量

本文核心解释变量源自于 Huang 和 Luk（2020）构建的贸易政策不确定性指数（Trade Policy Uncertainty，简记为 TPU）。该研究沿用 Baker 等（2016）创立的文本分析法，统计特定时段新闻报道中“贸易”、“政策”、“进出口”等关键词频率，通过标准化处理后得到 TPU 指数，能够全面反映中国企业生产经营中面临的国际经济环境不确定性。需要说明的是，前文所提到的 TPU 指数为月度数据，于是，本文在构造年度指标时，为更好模拟企业决策的实际情况，采用加权平均法计算年度 TPU 指数（*tpuweigh*）。具体做法为选取 4 月、8 月和 12 月月度指数，并根据企业决策过程中对环境因素的敏感程度，将权重分配为 1/6、1/3 和 1/2。为便于理解和解释，本文最后将加权平均后的数值除以 100，并滞后一期，得到本文基准回归部分所用的贸易政策不确定性指数（*L.tpuweigh*）。

（3）控制变量

在参考现有研究的基础上（Jory S R et al., 2020），本文选择控制企业财务特征、治理结构等层面的一系列变量，X 中主要包括企业年龄、资产负债率、总资产增长率、固定资产比率、股权集中度、总资产净利润率等。此外，该模型还控制行业固定效应和时间固定效应，用于剔除行业层面不随时间变化的因素及共同时间趋势对企业带来的影响。

（4）机制变量

行业竞争程度。本文采用赫芬达尔指数（*indus_hhi*）衡量行业竞争程度（谭庆美和魏东一，2014）。赫芬达尔指数采用各企业主营业务收入占行业内市场整体的主营业务收入比重的平方和计算，较高的赫芬达尔指数表明市场集中度较高、竞争程度较小，而较低的指数则表明市场集中度较小、行业竞争更为激烈。

绿色创新产出。本文以各年绿色专利申请总数（*patent*）来衡量该企业绿色创新产出能力（王永贵和李霞，2023），数据源于上市公司年报、上市公司社会

责任报告、上市公司网站信息、国家统计局、国家知识产权局、世界知识产权组织，并借助世界知识产权组织（WIPO）于2010年发布的环境友好型国际专利分类索引列表识别出企业绿色专利。

上述企业层面数据均来源于CSMAR数据库，具体变量设定见表3.1所列：

表3.1　主要变量描述

| 变量类型 | 变量名称 | 变量符号 | 测算方法 |
|---|---|---|---|
| **被解释变量** | 企业ESG表现 | *ESG* | 华证ESG评级数据 |
| **核心解释变量** | 贸易政策不确定性指标 | *TPU* | 贸易政策不确定性年度指数 |
| **控制变量** | 企业年龄 | *age* | ln(当年年份-企业上市年份) |
| | 资产负债率 | *lev* | 企业负债总计/企业资产总计 |
| | 总资产增长率 | *tagr* | 年末总资产的增长额/年初资产总额 |
| | 固定资产比率 | *far* | 固定资产/资产总额 |
| | 股权集中度 | *hindex* | 第一大股东持股比例平方和 |
| | 总资产净利润率 | *roa* | 净利润总额/企业资产平均总额 |
| | 行业竞争程度 | *indus_hhi* | 各企业主营业务收入占行业内市场整体的主营业务收入比重之平方和 |
| | 绿色创新产出 | *patent* | 企业当年绿色专利申请总数 |
| | 年度虚拟变量 | *Year FE* | 2010—2022年年度虚拟变量 |
| | 行业虚拟变量 | *Industry FE* | 按2012版证监会行业分类 |

## 3.3 描述性统计与相关性分析

表3.2呈现主要变量的描述性统计结果。企业ESG评级均值为4.03，最小值为1，最大值为8，反映出我国上市公司在绿色发展方面存在显著能力差异，整体ESG得分偏低。贸易政策不确定性指标则表现出较大的波动性。图3.1展示2010年与2020年行业HHI指数分布直方图，整体呈右偏分布。相较于2010年，2020年的更多行业HHI指数集中于1附近，表明整体行业竞争程度增加。同样，图3.2展示企业平均绿色专利申请数量的发展趋势，可见，平均绿色专利

申请书从 2010 年的 1.68 增长至 2018 年的 5.61，这一变化清晰展示出企业绿色创新的强劲动力和迅速增长态势。此外，本研究中其他变量的描述性统计数值均在合理区间内，样本企业之间存在较大差异性。

表 3.2　变量描述性统计

| 变量名称 | 变量符号 | 样本量 | 最小值 | 中位数 | 最大值 | 均值 | 标准差 |
|---|---|---|---|---|---|---|---|
| 企业 ESG 评级 | *ESG* | 18909 | 1 | 4 | 8 | 4.03 | 1.14 |
| 上期贸易政策不确定性指标 | L.*tpuweigh* | 18909 | 0.71 | 1.15 | 5.26 | 1.79 | 1.47 |
| 企业年龄 | *age* | 18909 | 0 | 14 | 30 | 13.33 | 6.6 |
| 资产负债率 | *lev* | 18909 | 0.05 | 0.48 | 0.98 | 0.47 | 0.22 |
| 总资产增长率 | *tagr* | 18909 | -0.37 | 0.08 | 3.29 | 0.17 | 0.43 |
| 固定资产比率 | *far* | 18909 | 0 | 0.19 | 0.73 | 0.23 | 0.18 |
| 股权集中度 | *hindex* | 18909 | 0.01 | 0.1 | 0.56 | 0.14 | 0.12 |
| 总资产净利润率 | *roa* | 18909 | -0.27 | 0.03 | 0.2 | 0.03 | 0.06 |
| 行业竞争程度 | *indus_hhi* | 18314 | 0 | 0.09 | 1 | 0.13 | 0.14 |
| 绿色创新 | *patent* | 14552 | 0 | 0 | 1543 | 3.64 | 37.17 |

表 3.3　相关系数矩阵

| 变量 | (1) | (2) | (3) | (4) | (5) | (6) | (7) | (8) | (9) |
|---|---|---|---|---|---|---|---|---|---|
| (1) *L.tpuweigh* | 1.000 | | | | | | | | |
| (2) *L.tpuari* | 0.975 | 1.000 | | | | | | | |
| (3) *L.tpugeo* | 0.977 | 0.998 | 1.000 | | | | | | |
| (4) *age* | 0.299 | 0.330 | 0.323 | 1.000 | | | | | |
| (5) *lev* | 0.012 | 0.013 | 0.013 | 0.260 | 1.000 | | | | |
| (6) *tagr* | -0.077 | -0.090 | -0.088 | -0.053 | 0.023 | 1.000 | | | |
| (7) *far* | -0.040 | -0.044 | -0.044 | -0.003 | 0.069 | -0.095 | 1.000 | | |
| (8) *hindex* | -0.054 | -0.061 | -0.059 | -0.038 | 0.072 | 0.059 | 0.073 | 1.000 | |
| (9) *roa* | -0.055 | -0.064 | -0.062 | -0.115 | -0.352 | 0.184 | -0.087 | 0.117 | 1.000 |

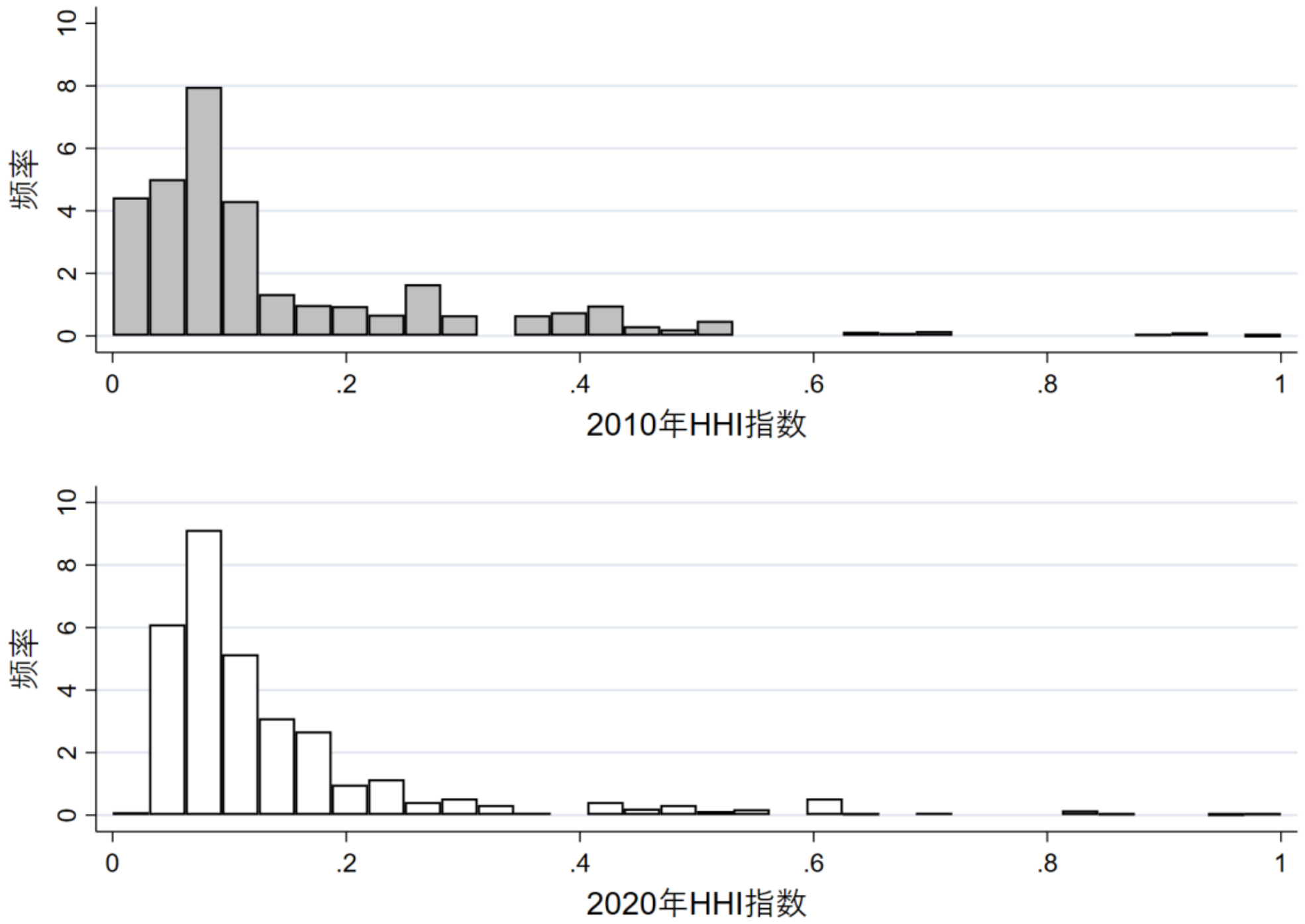

图 3.1　行业 HHI 指数分布

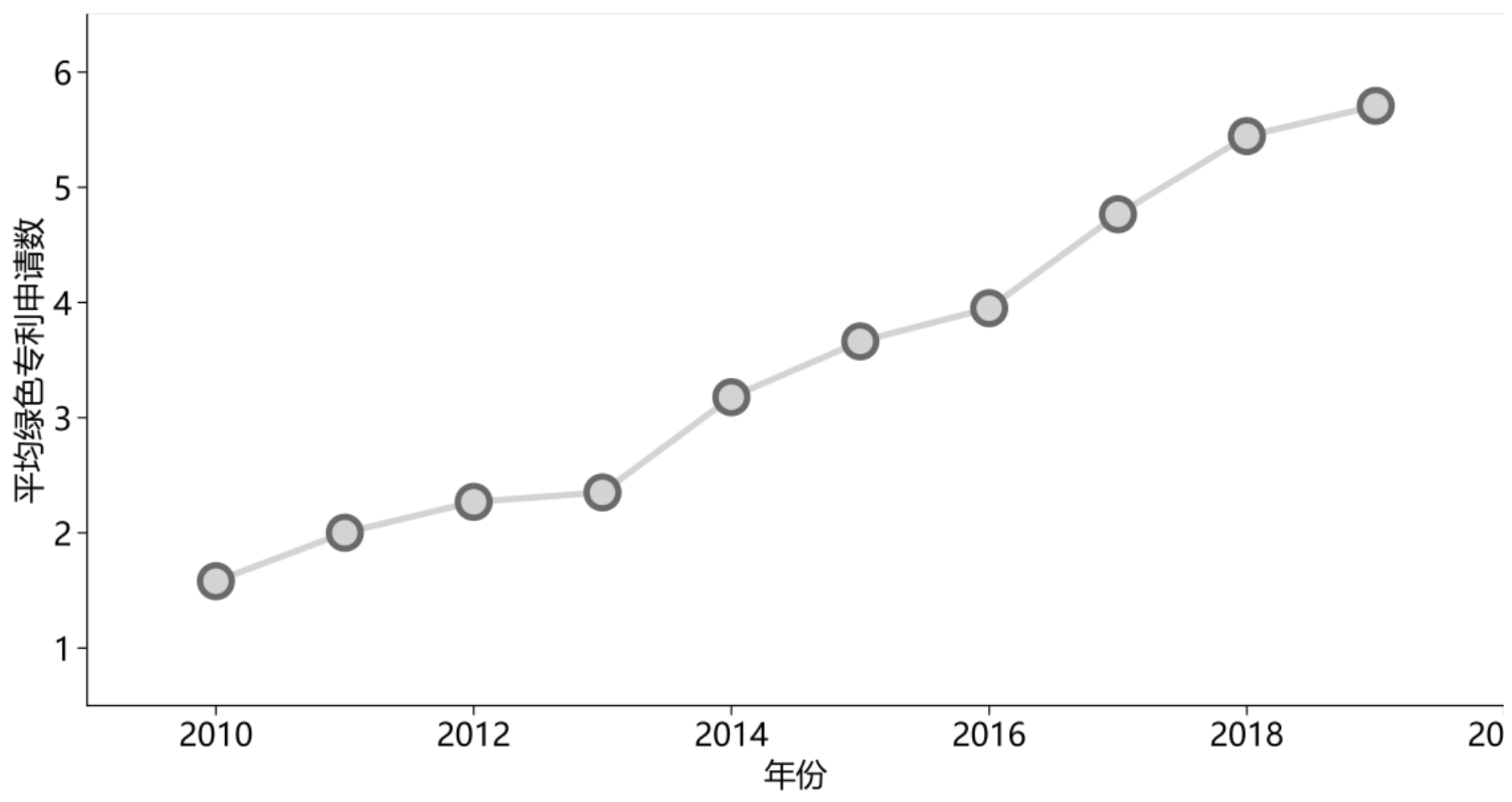

图 3.2　平均绿色专利申请数增长趋势

在进行实证分析前，本文还对主要变量进行相关性检验，结果如上表 4.3 所示。贸易政策不确定性与企业 ESG 表现呈显著正相关关系，初步判断贸易政策不确定性的负面影响可能会反向促进企业 ESG 表现的提升。除不同方法度量的核心解释变量（*L.tpuweigh*、*L.tpugeo* 和 *L.tpuari*）之间高度相关以外，其他变量

的相关系数均在 0.7 以下，表明各变量之间不存在严重的多重共线性。此外，本文还通过 Belsley 等人（1980）提供的 coldiag 2 方法来检验自变量之间是否存在多重共线性问题，若条件数（Condetion index）达到 30 或更高，则可能存在多重共线性问题。运行该命令后发现条件数为 13.49，小于临界值 30。综上所述，可以认为本文数据之间不存在明显的多重共线性问题。

# 第四章　实证结果分析

## 4.1 基准回归分析

表 4.1 报告关于贸易环境不确定性上升对企业 ESG 表现影响的基本估计结果。其中，第（1）列仅包含核心解释变量，而第（2）列则整合入全部控制变量，第（3）列在此基础上又加入年份固定效应和行业固定效应。无论是否考虑控制变量和固定效应，*L.tpuweigh* 系数均在 1%置信水平上显著为正，表明贸易政策不确定性能够提升企业绿色发展能力，促进 ESG 表现提升。

归纳众多已有研究，可将政策不确定性如何影响企业 ESG 表现的结论分为阻碍论和促进论两派。本文发现在一定程度上挑战了传统的阻碍论观点，即不确定性通常通常会负向影响企业发展，同时，也在印证了已有关于不确定性促进企业战略变革的相关结论。进一步分析认为，贸易政策不确定性导致的倒逼效应，实质上是在促使企业将 ESG 表现视作为一项关键的战略工具，这种策略转变有两个主要动因：一方面，提升 ESG 表现能够向利益相关者传递一系列积极信号，进而吸引更多注重可持续发展和社会责任的投资者，为企业带来广泛的市场关注，此举不但能够提升企业形象，还能降低资金成本，并吸引更多融资机会；另一方面，鉴于全球市场对企业社会责任和环境影响的日益关注，良好的 ESG 表现已是保持竞争力的关键因素。因此，企业通过提高其 ESG 表现，不仅有助于体现其绿色发展承诺，还能在一定程度上减少因贸易政策波动而引发的负面影响。

从控制变量看，首先，企业年龄越小，上市企业 ESG 表现更佳。年轻企业更倾向于采纳新兴商业模式和管理理念，易于适应市场及社会对绿色发展的需求；其次，总资产增长率（*tagr*）较低时，上市企业 ESG 表现更优。增长缓慢或

者更加注重内部质量而非外部扩张的企业，会更多关注长期、可持续性投资；第三，股权集中度（*hindex*）越高，上市企业 ESG 表现越好。这表明少数股东拥有更打控制权，公司便会更重视长期利益而非短期回报；最后，资产回报率（*roa*）较高的上市企业 ESG 表现也更优。企业经营效率越高，盈利能力越强，就有越能力承担 ESG 投资的初期成本，同时，经济表现良好的企业会受到更多公众关注，也就更有动力提高 ESG 表现以塑造良好的公众形象。

表 4.1　基准回归

| | （1） | （2） | （3） |
|---|---|---|---|
| *L.tpuweigh* | 0.0235* | 0.0464*** | 0.173*** |
| | (0.0108) | (0.0111) | (0.0168) |
| *lnage* | | -0.105*** | -0.319*** |
| | | (0.0200) | (0.0235) |
| *far* | | -0.542*** | 0.481*** |
| | | (0.0777) | (0.0928) |
| *tagr* | | -0.288*** | -0.261*** |
| | | (0.0322) | (0.0328) |
| *lev* | | 0.235** | -0.0570 |
| | | (0.0745) | (0.0793) |
| *hindex* | | 1.469*** | 1.312*** |
| | | (0.116) | (0.119) |
| *roa* | | 6.055*** | 6.442*** |
| | | (0.272) | (0.277) |
| *cut1* | -3.545*** | -3.533*** | -2.930*** |
| | (0.0465) | (0.0690) | (0.128) |
| *cut2* | -2.294*** | -2.258*** | -1.640*** |
| | (0.0298) | (0.0593) | (0.124) |
| *cut3* | -0.871*** | -0.784*** | -0.118 |
| | (0.0230) | (0.0573) | (0.123) |
| *cut4* | 0.685*** | 0.832*** | 1.591*** |
| | (0.0227) | (0.0578) | (0.123) |
| *cut5* | 2.485*** | 2.673*** | 3.521*** |
| | (0.0304) | (0.0601) | (0.125) |
| *cut6* | 5.211*** | 5.413*** | 6.293*** |
| | (0.0965) | (0.109) | (0.154) |
| *cut7* | 8.098*** | 8.301*** | 9.184*** |
| | (0.408) | (0.412) | (0.426) |
| *Year FE* | NO | NO | YES |
| *Industry FE* | NO | NO | YES |
| N | 18909 | 18908 | 18908 |
| Wald | 4.70*** | 970.50*** | 2550.89*** |

注：*、**、***分别表示 10%、5%、1%显著性水平，括号内为异方差稳健标准误，下同。

由于基准回归部分使用的有序 Logit 回归模型，其回归系数无法直接反映贸易政策不确定性对企业 ESG 表现的影响程度，于是本文在此基础上进行边际效应分析，结果如表 4.2 所示：当贸易政策不确定程度每增加 1，企业 ESG 表现为 5 的概率增加约 2.34%，ESG 表现为 6 的概率增加 1.12%，ESG 表现为 7 的概率增加 0.09%。这些边际效应数据提供了一个更具体的理解，即贸易政策不确定性对企业 ESG 表现的不同层次产生不同幅度的影响。

表 4.2 边际效应分析

| 变量 | ESG=1 | ESG=2 | ESG=3 | ESG=4 |
|---|---|---|---|---|
| *L.tpuweigh* | -0.0044*** | -0.0088*** | -0.0192*** | -0.0032*** |
| | (0.0005) | (0.0009) | (0.0019) | (0.0005) |
| *Control* | YES | YES | YES | YES |
| 变量 | ESG=5 | ESG=6 | ESG=7 | ESG=8 |
| *L.tpuweigh* | 0.02338*** | 0.0112*** | 0.0009*** | 0.0001* |
| | (0.0023) | (0.0012) | (0.0001) | (0.0000) |
| *Control* | YES | YES | YES | YES |

## 4.2 稳健性检验

为确保研究结果的稳健性和可信度，本研究进行了五种不同的稳健性检验：更换被解释变量的度量方法、重新测算核心解释变量、剔除时间维度和空间维度的部分样本、模型替换、提高聚类标准以及控制其他政策不确定性。接下来将详细介绍这些稳健性检验的方法和结果。

### 4.2.1 更换被解释变量的度量方法

为进一步验证主回归结论，本文采用同样被学界广泛认可和使用的彭博 ESG 评级数据（*ESG_bb*）对基准回归使用的衡量指标进行替换。不同于基准回归仅采用年末评级数据，本节引入华证年度内四次季度评级的平均值（*ESG_ssi_avg*）作为新被解释变量。结果见表 4.3 第（1）、（2）列，在替换被解释变量后，贸易政策不确定性上升仍显著提升企业 ESG 表现，此发现能够有效排除对第三方评级机构在 ESG 评分上可能存在偏见的疑虑

### 4.2.2 重新测算核心解释变量

在基准回归中年度指标由加权平均法构造而来，出于稳健性考量，此处对核心解释变量的度量方法进行替换。如表 4.3 第（3）、（4）列所示，在分别计算月度指标的几何平均值（*L.tpuari*）和算术平均值（*L.tpugro*）代入回归方程后，系数均在 1%水平下显著为正，回归结果高度稳健。

表 4.3　替换变量的稳健性检验

| | （1） | （2） | （3） | （4） |
|---|---|---|---|---|
| | ESG_bb | ESG_ssi_avg | ESG | ESG |
| *L.tpuweigh* | $1.428^{***}$ | $0.108^{***}$ | | |
| | (0.0993) | (0.00935) | | |
| *L.tpuari* | | | $0.198^{***}$ | |
| | | | (0.0192) | |
| *L.tpugro* | | | | $0.229^{***}$ |
| | | | | (0.0222) |
| *lnage* | 0.929*** | -0.00742 | -2.951*** | -2.932*** |
| | (0.168) | (0.0155) | (0.128) | (0.128) |
| *far* | 2.449*** | 0.336*** | -1.662*** | -1.642*** |
| | (0.541) | (0.0528) | (0.123) | (0.123) |
| *tagr* | -0.921*** | -0.117*** | -0.140 | -0.120 |
| | (0.231) | (0.0195) | (0.122) | (0.123) |
| *lev* | 4.869*** | -0.138** | 1.570*** | 1.589*** |
| | (0.468) | (0.0450) | (0.123) | (0.123) |
| *hindex* | 3.855*** | 0.690*** | 3.499*** | 3.519*** |
| | (0.678) | (0.0678) | (0.125) | (0.125) |
| *roa* | 12.61*** | 3.320*** | 6.271*** | 6.290*** |
| | (1.377) | (0.153) | (0.154) | (0.154) |
| *_cons* | 6.755*** | 2.961*** | 9.162*** | 9.181*** |
| | (0.641) | (0.0786) | (0.426) | (0.426) |
| *Year FE* | YES | YES | YES | YES |
| *Industry FE* | YES | YES | YES | YES |
| N | 7049 | 18434 | 18908 | 18908 |
| $R^2$/Wald | 0.166 | 0.125 | 2550.89*** | 2550.89*** |

### 4.2.3 剔除时间维度和空间维度的部分样本

在本研究中，基准模型的回归样本涵盖了 2015 年中国股市异常波动时期，这一时期对金融市场投资者的投资意愿产生较大抑制作用。而推动企业进行 ESG 工作的重要推力之一就是投资者的期望，投资者对 ESG 的关注力度和态度，能够从很大程度上推动和督促上市公司更好地在 ESG 方面取得成绩。因此，为削弱金融市场因素干扰，本文将 2015 年样本剔除后重新进行回归，结果见表 4.4 第（1）列，最终得到与基准模型一致的结果。同时，美国进口贸易结构变动对中国制造业发展的影响程度最深，考虑到这一因素影响，本文剔除制造业企业，重新对基准模型进行估计，结果列于表 4.8 第（2）列，也与基准模型一致。

### 4.2.4 模型替换

出于结论稳健性考量，本文将原有序 Logit 模型重新替换为“最小二乘虚拟变量法”（Least Square Dummy Variable，简称 LSDV）进行回归。结果如表 4.4 第（3）列所示，显示结果依然稳健，说明贸易政策不确定性显著促进企业 ESG 表现。

### 4.2.5 控制其他政策不确定性

为增强回归结果稳健性，排除其他政策不确定性干扰，本文在基准回归中引入由 Huang 和 Luk（2020）构建的财政政策不确定性（*FPU*）、汇率政策不确定性（*EXPU*）与货币政策不确定性（*MPU*）[2]。据表 4.4 第（4）列回归结果显示，在控制其他政策不确定性后，核心解释变量的方向与显著性与基准回归结果大致相符，此结果进一步验证了基准回归结果的稳健性。

表 4.4　稳健性检验

| | (1) | (2) | (3) | (4) |
|---|---|---|---|---|
| | 剔除 2015 年 | 剔除制造业 | LSDV | 控制其他政策不确定性 |
| *L.tpuweigh* | 0.173*** | 0.187*** | 0.0924*** | 0.361** |
| | (0.0168) | (0.0254) | (0.00971) | (0.124) |
| *L.fpuweigh* | | | | -0.0115 |
| | | | | (0.00610) |
| *L.expuweigh* | | | | -0.0192*** |
| | | | | (0.00484) |
| *L.mpuweigh* | | | | 0.0482** |
| | | | | (0.0187) |
| *lnage* | -0.322*** | -0.400*** | -0.183*** | -0.319*** |
| | (0.0241) | (0.0403) | (0.0139) | (0.0235) |
| *far* | 0.518*** | 0.352* | 0.251*** | 0.481*** |

[2] 上述年度政策不确定性指数均通过 3.2 节所述之加权平均法计算，并经滞后一期处理后加入回归。

| | | | | |
|---|---|---|---|---|
| | (0.0971) | (0.138) | (0.0547) | (0.0928) |
| *tagr* | -0.253*** | -0.317*** | -0.146*** | -0.261*** |
| | (0.0352) | (0.0469) | (0.0202) | (0.0328) |
| *lev* | -0.0853 | 0.0458 | -0.103* | -0.0570 |
| | (0.0835) | (0.120) | (0.0471) | (0.0793) |
| *hindex* | 1.314*** | 1.847*** | 0.840*** | 1.312*** |
| | (0.124) | (0.173) | (0.0700) | (0.119) |
| *roa* | 6.338*** | 5.761*** | 3.569*** | 6.442*** |
| | (0.288) | (0.449) | (0.161) | (0.277) |
| *cut1* | -2.942*** | -3.001*** | | -1.855* |
| | (0.133) | (0.166) | | (0.877) |
| *cut2* | -1.646*** | -1.668*** | | -0.565 |
| | (0.127) | (0.154) | | (0.876) |
| *cut3* | -0.119 | -0.166 | | 0.957 |
| | (0.126) | (0.153) | | (0.876) |
| *cut4* | 1.594*** | 1.562*** | | 2.666** |
| | (0.127) | (0.154) | | (0.876) |
| *cut5* | 3.514*** | 3.458*** | | 4.596*** |
| | (0.129) | (0.157) | | (0.876) |
| *cut6* | 6.224*** | 6.250*** | | 7.368*** |
| | (0.157) | (0.198) | | (0.881) |
| *cut7* | 9.107*** | 9.225*** | | 10.26*** |
| | (0.427) | (0.598) | | (0.974) |
| _cons | | | 3.562*** | |
| | | | (0.0760) | |
| *Year FE* | YES | YES | YES | YES |
| *Industry FE* | YES | YES | YES | YES |
| N | 17189 | 7765 | 18908 | 18908 |
| $R^2$/Wald | 2366.34*** | 1707.94*** | 0.119 | 2550.89*** |

## 4.3 内生性处理

当前理论并不支持企业 ESG 表现增强会导致贸易政策不确定性上升这一观点，故本研究对反向因果关系存在的可能性抱有怀疑。一方面，贸易政策的不确定性波动主要源于宏观层面，如政府政策更迭、国际经济动荡及国际政治因素等，这些宏观变化直接影响企业运营环境，进而作用于企业 ESG 表现；另一方面，微观企业的行为和决策对贸易政策不确定性的影响相对有限，因为后者的形成是多种复杂因素交织的结果，不易受单个企业行为的直接影响。为解决内生性问题，本文已在初期的数据处理和计量模型构建方面对核心解释变量进行滞后一期处理，这一处理方式能够在一定程度上降低反向因果关系的风险，并缓解内生性问题。除此之外，为避免其他潜在的内生性问题对回归估计结果的干扰，本文将进行工具变量法、倾向得分匹配法（PSM）和安慰剂测试。

### 4.3.1 工具变量法

为增强稳健性，本文于此处将工具变量法和 CMP 估计法（Conditional Mixed Process, 条件混合过程）相结合以进一步处理潜在的反向因果问题。首先，在工具变量选取方面，考虑到美国和日本分别是中国的第一和第二大贸易伙伴，其内部贸易政策不确定性很容易对我国产生影响，但又不会直接影响我国企业行为。于是，本文引入与中国存在贸易联系的其他国家的贸易政策不确定性指数。接着，本文利用 Caldara 等（2020）在政策不确定性指数网站[3]公开提供的美国与日本 TPU 指标，根据中美与中日贸易总额对这两国 TPU 指数进行加权平均，并得到最终工具变量（*L.usjp_tpuweigh*）。最后，本文选取上述 TPU 指数的滞后一期作为中国贸易政策不确定性的工具变量。

需要注意的是，本文于基准回归部分采用有序 Logit 模型以估计贸易政策不确定性对企业 ESG 表现的影响，但从技术可行角度来看，无法直接在排序模型之基础上使用工具变量法。因此，对于包含内生变量的有序 Logit 模型，将工具变量法和 CMP 估计法相结合以更好解决模型内生性问题（Roodman, 2011）。又由于 CMP 估计无法进行工具变量有效性检验，采用线性模型两阶段最小二乘估

[3] 美国和日本 TPU 指数来源网站：https://www.matteoiacoviello.com/tpu.htm#data

计的弱工具变量检验方法来进行工具变量有效性检验（Chyi & Mao, 2012）。结果显示，Kleibergen-Paap rk Wald F 第一阶段统计量估计结果的 F 值远大于 10% maximal IV size 所对应的临界值 16.38，表明本文所选工具变量均不存在弱相关问题。此外，Kleibergen-Paap rk LM 的 p 值在 1%水平上显著，证明不存在工具变量不可识别问题。

表 4.5 中第（1）列为基于美日加权 TPU 指数作为工具变量的 CMP 估计结果，结果显示，贸易政策不确定性对企业 ESG 表现的影响系数与基准回归结果非常接近。说明在使用 CMP 估计处理内生性问题后，贸易政策不确定性促进企业 ESG 表现提升的结论依然成立。

### 4.3.2 倾向得分匹配法（PSM）

本文估计结果可能会因为样本自选择偏误而引致内生性问题。具体来看，注重 ESG 表现的企业进一步提高 ESG 水平所需成本相对较低。反之，ESG 表现不佳的企业在提升该领域表现时，面临更高投入的成本。因此，ESG 表现不同的企业在提升 ESG 表现方面的动因存在显著差异。

为消除样本自选择所致内生性问题，本研究采用倾向得分匹配（PSM）方法进行处理。以样本中 50%的 ESG 表现得分作为分界线，将 ESG 表现得分低于此界限的企业归为控制组，其余为实验组。按照企业特征变量进行 1:1 最邻近匹配，协变量包括资产负债率、总资产增长率、固定资产比率、股权集中度和总资产净利润率。图 4.1 为采用有序 Logit 模型对样本进行 1:1 匹配后的结果。匹配后，两组样本的标准偏误绝对值均低于 10%，无法拒绝处理组与对照组间无显著差异的原假设，表明匹配后两组样本在变量特征上趋于一致，匹配结果达到平衡。接下来，对匹配后样本进行回归分析，相应结果见表 4.5 第（2）列。回归结果显示，主要解释变量系数在 1%水平上显著为正，与基准回归一致，说明本研究所用样本不存在严重的自选择问题。

需要特别说明的是，PSM 回归分析中所采用的替代被解释变量为华证 ESG 季度评级均值，依据上文稳健性检验结果，该指标作为 ESG 表现的代理变量具有高度稳健性，故而可以将其视为一个合理的代理变量。鉴于基准回归采用的华证 ESG 年末评级数据为 1 至 9 分的离散变量，与之对比，稳健性检验中采用的华证 ESG 季度评级均值为连续变量，这一性质使其更适用于作为分组依据。因

此，在 PSM 回归分析中，本文决定采用表现稳健的华证 ESG 季度评级均值作为 ESG 代理变量。

### 4.3.3 安慰剂测试

为缓解观察结果可能会受某些无法观测的时变特征影响的问题，进行安慰剂测试。在测试中，对主要解释变量（*L.tpuweigh*）在企业 ESG 评级当年进行随机赋值，此种随机化方法的优势在于其能在保持数据结构中的时变特征不变的同时，进行有效的分析。若企业 ESG 表现的提升主要由贸易政策不确定性所驱动，则基准回归结果应该较安慰剂测试所得估计更强。表 4.5 中第（3）列呈现的主要解释变量随机处理后的回归结果显示，*L.tpuweigh* 平均系数为 0.0304，远小于上文表 4.1 第（3）列中的系数 0.173。这表明本研究估计结果受到不可观测因素影响的可能性较小。进而，可以认为，企业 ESG 表现的改善并非由随机性因素驱动，而是受到贸易政策不确定性的正向影响。

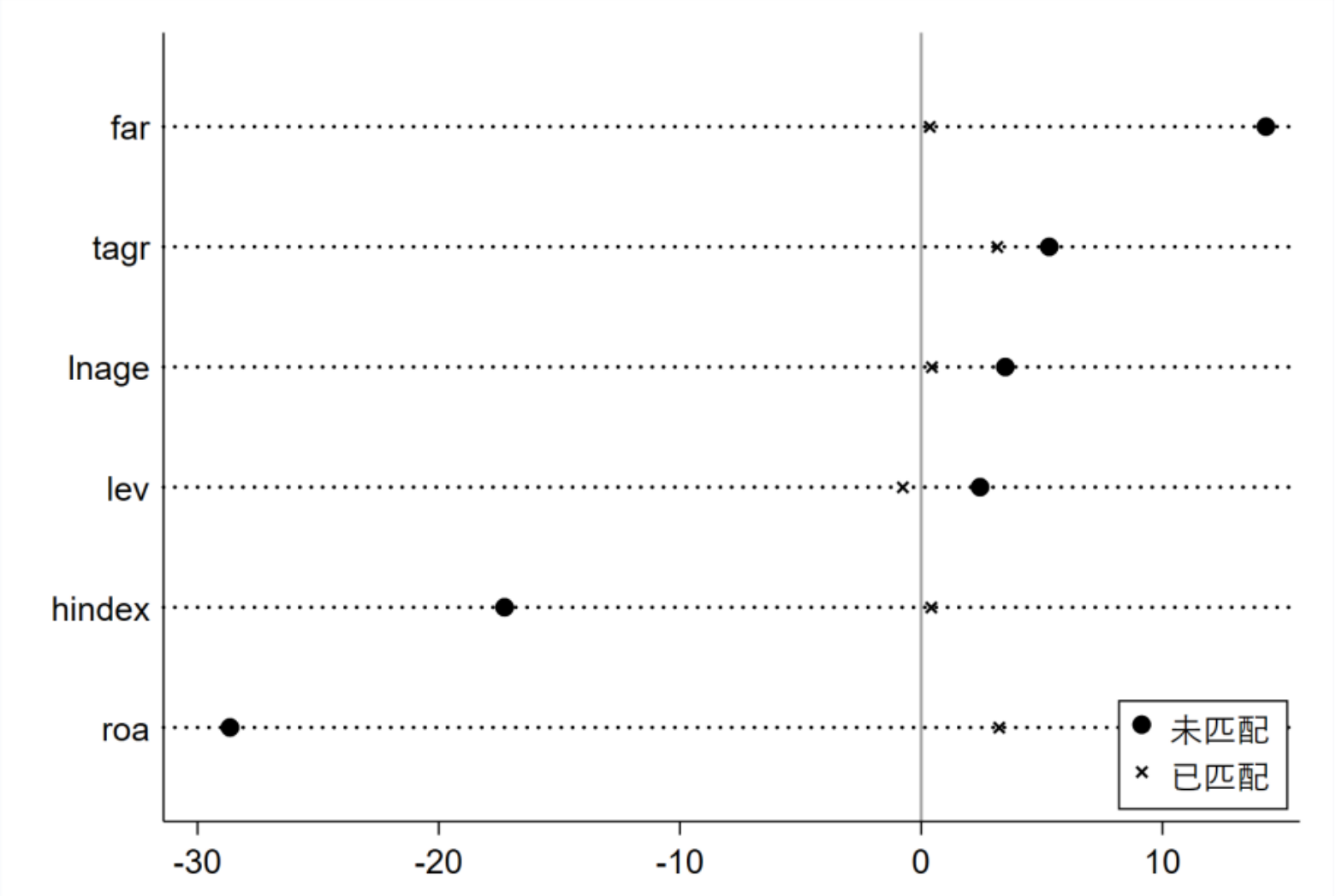

图 4.1　协变量间的标准化%偏差

表 4.5　内生性检验

| | （1） | （2） | （3） |
|---|---|---|---|
| | ESG | ESG_ssi_avg | ESG |
| *L.tpuweigh* | 0.037 | 0.104*** | 0.0304*** |
| | (0.0054) | (0.0265) | (0.00644) |
| *lnage* | -0.111 | 0.263*** | -0.119*** |
| | (0.0132) | (0.0402) | (0.0124) |
| *far* | 0.209 | 0.385** | 0.214*** |
| | (0.0510) | (0.139) | (0.0541) |
| *tagr* | -0.155 | 0.044 | -0.159*** |
| | (0.0181) | (0.0600) | (0.0205) |
| *lev* | -0.128 | -0.191 | -0.156** |
| | (0.0419) | (0.121) | (0.0489) |
| *hindex* | 0.797 | 0.074 | 0.830*** |
| | (0.0674) | (0.171) | (0.0709) |
| *roa* | 3.326 | 2.054*** | 3.442*** |
| | (0.1354) | (0.480) | (0.158) |
| _cons | | | 3.743*** |
| | | | (0.0736) |
| *Year FE* | YES | YES | YES |
| *Industry FE* | YES | YES | YES |
| N | 18908 | 8193 | 18908 |
| $R^2$/Wald | 2274.54*** | 211.56*** | 0.112 |

# 第五章　异质性检验与机制分析

## 5.1 异质性检验

在进行分组回归检验异质性之前，本文先对变量进行中心化处理，以消除共线性和个体尺度差异的影响。此外，为更全面地探究组间系数差异，本文使用基于似无相关模型 SUR 的检验（Seemingly Unrelated Estimation Test, SUEST）进行估计，此种方法统计效力较高，能够更精确判定组间系数差异。

### 5.1.1 企业技术特征

技术是实现企业可持续发展目标的关键要素。由于高新技术企业与其他企业相比，在应对贸易政策不确定性以及制定自身的 ESG 表现方面存在显著差异，

所以其应对贸易政策不确定性的策略和自身 ESG 表现也各不相同。基于此，本文参照《高新技术企业认定管理办法》和《国家重点支持的高新技术领域》，将企业样本划分为高新技术企业和非高新技术企业。通过表 5.1 的第（1）、（2）列可以看出，高新技术组核心解释变量回归系数大于非高新技术组，这初步表明，贸易政策不确定性带来的企业 ESG 表现提升作用在高新技术企业中更为显著。接着对两组核心解释变量的回归系数进行 SUEST 检验，得到 chi2 (1) =13.05，Prob>chi2=0.0003，表明贸易政策不确定性对企业 ESG 表现的推动作用受到企业技术特征异质性的影响。

这些企业往往在技术创新、研发投入和市场扩张方面具有高度依赖性，且频繁参与国际合作和贸易活动。因此，贸易政策的不确定性会直接影响这些企业的国际市场准入、供应链稳定性和技术合作机会。但是，高新技术企业也具备更强的技术适应性，能够快速整合、构建及重新配置内外资源和能力，以适应不确定的贸易政策环境。同时，高新技术企业倾向于加大对研发和创新活动的投入，相较于其他企业类型，其创新能力表现更为显著。强化的创新能力显著提高了高新技术企业的存活概率，从而在贸易政策波动环境下减缓了其所承受的负面冲击（Dongyang Zhang et al., 2018）。于是，高新技术企业能够在应对贸易政策不确定性时表现出更高的适应性和韧性，赢得更佳的 ESG 表现。

### 5.1.2 企业内部控制

内部控制的实质是风险控制，在企业运营和决策过程中扮演关键角色，其实施程度在很大程度上决定着企业应对外部不确定性的能力。本文根据年度企业内部控制指数中值，将样本划分为低内控组和高内控组，结果如表 5.1 中列（3）、（4）所示。高内控组中，核心解释变量系数在 1%水平上显著为正，而低内控组该系数统计上不显著。这表明在贸易政策不确定性加剧下，内部控制更严格的企业更倾向于提升 ESG 表现。为进一步验证这一发现，本研究对两组核心解释变量回归系数进行 SUEST 检验，结果显示：chi2 (1) = 36.87，Prob > chi2 = 0.0000，从而证明贸易政策不确定性对企业 ESG 表现的推动作用受内部控制程度差异显著影响。经济政策不确定性会促进企业战略变革，而良好的内部控制体系能够为战略制定提供可靠的信息和决策支持。因此，本文可以推断，在贸易政策不确定性环境下，内部控制程度较高的企业更倾向于部署 ESG 战略，推动 ESG 变革。

又由综述内容揭示，贸易政策不确定性增强企业财务压力，进而对其 ESG 表现提升构成阻碍。然而，在这一部分可以观察到，在贸易逆境中，仅有内部控制严格的企业才能有效提升 ESG 表现。基于此，本文推断，可能原因在于企业内部控制质量越高，企业面临的银行债务融资成本越低。同理，在贸易政策不确定性的环境下，内部控制水平高的企业同样能够获得较低的融资成本，这种财务灵活性能够为企业提升 ESG 表现提供必要的资源和能力，从而缓解企业融资压力。

### 5.1.3 CEO 环保背景

在波动性贸易环境中，企业 ESG 表现提升也可能受到高管环保背景的影响。本文将样本企业按 CEO 是否具有环保背景分为两类，进行回归分析。如表 5.1 所示，列（5）和列（6）分别对应无环保背景与有环保背景的企业。两组样本中，核心解释变量系数均达到 1%显著水平，但环保背景组中核心解释变量系数显著高于前者。进一步，通过 SUEST 检验对核心解释变量进行比较，结果显示 chi2 (1) = 9.51，Prob > chi2 = 0.0020，表明两组间差异显著。显而易见，在贸易政策不确定性上升的现状下，拥有环保背景 CEO 领导的企业在提升 ESG 表现方面展现出更强烈的意愿。

这一发现进一步支持高层梯队理论的观点，即管理者的个人特质和背景对企业策略和行为具有显著影响，尤其是在贸易政策不确定性的背景下，这一影响变得更加显著。首先，依据高层梯队理论，管理者特质能显著影响其战略选择及行为倾向，在这个框架下，CEO 的个人属性，如价值观、经验和教育背景，被视为影响其决策和行为的关键因素（Bass Bernard M, 1999）。作为决策核心的 CEO 若同时具备环保背景，便能够对其他高管施加压力，将其自身对环境问题的关注融入整个高管团队。因此，具备环保背景的 CEO 在推动企业专注于绿色发展和提升 ESG 表现方面扮演着至关重要的角色。接着，面对贸易政策不确定性带来的市场压力，高管团队的环保意识变得尤为重要。这种意识不仅有助于将市场里转化为潜在的绿色发展市场机遇，而且还能鼓励企业采纳和实施绿色战略。这样的战略转变使企业能够在竞争激烈的市场环境中脱颖而出，甚至可能在挑战中找到新的增长点（曹洪军和陈泽文，2017）。最后，考虑到 ESG 表现是绿色投资决策的关键考量因素，高管团队的环保背景和承诺在此环节显得尤为重要。相应

研究揭示，高管具备环保背景及相关从业经验，能够显著提升企业履行环境责任的能力（李毅，2023）。这种态度不仅能够提高企业吸引绿色投资者的潜力，而且能为企业在市场上建立生态友好型品牌形象、吸引生态友好型资本创造有利条件。

表 5.1　异质性检验

| | 企业类型 | | 内部控制力度 | | CEO 环保背景 | |
|---|---|---|---|---|---|---|
| | （1） | （2） | （3） | （4） | （5） | （6） |
| | 高新技术 | 非高新技术 | 低内控 | 高内控 | 无环保背景 | 有环保背景 |
| *L.tpuweigh* | 0.442*** | 0.190*** | -0.0359 | 0.286*** | 0.194*** | 0.472*** |
| | (0.0648) | (0.0240) | (0.0450) | (0.0258) | (0.0232) | (0.0825) |
| *lnage* | -1.021*** | -0.447*** | -0.311* | -0.478*** | -0.461*** | -0.910*** |
| | (0.167) | (0.0807) | (0.149) | (0.0813) | (0.0739) | (0.236) |
| *far* | -0.0112 | -0.305 | -0.258 | -0.317 | -0.384* | 0.906 |
| | (0.514) | (0.182) | (0.296) | (0.210) | (0.180) | (0.585) |
| *tagr* | 0.165 | -0.0853* | 0.0580 | -0.110** | -0.0807* | 0.138 |
| | (0.104) | (0.0348) | (0.0626) | (0.0390) | (0.0348) | (0.107) |
| *leverage* | -0.798* | -1.221*** | -1.677*** | -0.862*** | -1.214*** | -1.075* |
| | (0.383) | (0.138) | (0.216) | (0.162) | (0.135) | (0.463) |
| *hindex* | -1.225 | 0.727** | 1.806*** | -0.122 | 0.566* | -0.440 |
| | (0.863) | (0.280) | (0.521) | (0.310) | (0.276) | (1.020) |
| *roa* | 1.487 | 1.683*** | -0.424 | 2.485*** | 1.622*** | 1.885 |
| | (0.829) | (0.306) | (0.429) | (0.389) | (0.296) | (1.141) |
| *Year FE* | YES | YES | YES | YES | YES | YES |
| *Industry FE* | YES | YES | YES | YES | YES | YES |
| N | 2904 | 16004 | 5220 | 13688 | 17297 | 1611 |
| Wald | 121.29*** | 385.68*** | 171.91*** | 409.37*** | 129.07*** | 392.65*** |

## 5.2 机制分析

前文已证实贸易政策不确定性在提升企业 ESG 表现方面发挥显著促进作用。那么这一现象的背后机制究竟如何？本文认为，行业竞争和绿色创新可能是关键的影响渠道。于是，为了深入探索并验证贸易政策不确定性如何通过这些渠道促进企业 ESG 表现，本文采用中介效应检验方法，对前文理论分析部分提出的贸易政策不确定性影响企业 ESG 表现的作用机制进行研究。参考江艇（2022）对因果推断研究中的中介效应分析建议，本文采用“两步法”来检验中介效应，此法重点关注贸易政策不确定性对中介变量的影响，用于解释因果关系较为清晰直观。在基准回归模型 4.2 的基础上，设定模型 5.1 进行影响机制检验：

$$M_t = \alpha + \beta_1 L.TPU + X'\Gamma + \omega_t + \eta_i + \varepsilon_{it} \tag{5.1}$$

在本研究中，M 代表企业所处行业竞争程度和绿色创新。本文采用基于主营业务收入计算的赫芬达尔-赫希曼指数（*Indus_hhi*）来衡量行业竞争程度。以绿色专利申请数量衡量企业的绿色创新产出能力（*patent*）。

相较于其他中介效应检验方法，Bootstrap 方法因其较高的统计效力而被广泛认可，能够有效取代 Sobel 方法，直接对系数乘积进行检验（温忠麟和叶宝娟，2014）。因此，为进一步验证本文研究结果，本研究通过 Bootstrap 检验方法，对原始数据进行 1000 次有放回抽样，根据结果判断中介效应是否存在，以增强机制检验的完备性和可信度。

### 5.2.1 行业竞争效应

基准回归的结果支持战略性增长期权理论，上文提到，该理论的基本假设之一就是产品市场为不完全竞争市场，存在较为激烈的差异化竞争。ESG 作为长期主义和价值投资，企业提升 ESG 表现的目的就是提升自身的竞争优势以获取更多的市场份额。因此，战略性增长期权理论框架中行业竞争会影响企业在不确定性环境下的 ESG 表现。

表 5.2 第（1）列为行业竞争渠道的验证过程。第（1）列中，核心解释变量系数在 1%水平上显著且为负，说明贸易政策不确定性会加剧企业所处行业的竞争，这与前文理论分析部分结论相呼应。处于稳健考量，本文进一步通过 Bootstrap 法对中介机制进行检验，结果显示：间接（中介）效应_bs_1 的 p 值为 0.00。同时，行业竞争效应的百分位及偏差校正后的置信区间分别为[ 0.0026 , 0.0007 ]和

[ 0.0014 , 0.0041 ]，均未包含 0，表明贸易政策不确定性通过行业竞争渠道影响企业 ESG 表现的论断成立。

本文认为，从市场重心转移的视角来看，贸易政策波动导致行业竞争程度加剧的现象便可以得到合理解释。贸易政策不确定性，常常伴随着关税波动、供应链中断以及市场准入规则改变，会首先对企业的国际贸易行为产生直接影响。例如，出口导向型企业可能会因为关税上升而在海外市场份额上受损，而依赖进口的企业则可能因供应链的断裂而面临原材料短缺的困境。这些外部冲击都在推动企业全球市场上重新布局。另外，贸易政策不确定性也会影响企业的投资决策和长期规划。在面临不稳定的贸易环境时，企业可能会缩减跨境投资和扩张计划，而更多地依赖国内市场。这种战略上的转变会导致更多企业将注意力集中在同一国内市场上，从而加剧市场竞争压力。

在如此激烈的竞争背景下，企业对产品市场竞争压力会更加敏感，寻求新的竞争优势成为企业的必然选择。于是，进一步，在高度竞争的行业环境中，贸易政策不确定性就会对企业的 ESG 表现产生显著的“倒逼”作用。因为在竞争日益激烈的市场环境中，这种不确定性不仅是巨大挑战之一，而且将成为推动企业更加关注 ESG 表现的关键动力。如此，在贸易政策不确定性导致的激烈竞争环境中，企业 ESG 表现的提升就会成为获取竞争优势的关键途径。

### 5.2.2 绿色创新效应

表 5.2 第（2）列为绿色创新渠道的验证过程，其中，核心解释变量系数在 1%水平下显著为正，说明当企业面临来自贸易政策不确定性的冲击时，挑战与机遇并存，企业选择抓住机遇迎难而上，提升绿色技术创新产出，从而巩固企业所处的市场地位。进一步，出于稳健性考虑，本文通过 Bootstrap 法进行中介机制检验，结果显示：间接（中介）效应_bs_1 的 p 值为 0.00。同时，绿色创新效应的百分位及偏差校正后的置信区间分别为[ 0.0015 , 0.0052 ]和[ 0.1015 , 0.1536 ]，均不包含 0。表明贸易政策不确定性影响企业 ESG 表现的绿色创新渠道成立，也印证了理论分析中关于该机制的初步推断。在理论分析部分，本文已对贸易政策不确定性激励企业进行绿色创新的观点论点进行了阐述——贸易政策不确定性为企业提供一种特殊的外部经济风险，绿色创新则成为企业响应这一

不确定性、实现可持续发展的重要路径。接下来，可先从绿色创新定义和现实影响两方面出发，以理解绿色创新促进企业 ESG 表现提升的背后原因：

一方面，根据目的不同，绿色创新可被分为以下三类：I类绿色创新，主要是为了节约资源，间接带来减少碳排放等提高环境绩效的结果；II类绿色创新，即专门为减少环境破坏而进行的创新；III类绿色创新，是通过合理设计环境规制引发的创新。结合绿色创新的定义可以看出，无论是何种绿色创新，最终都会给环境保护带来正面影响。

另一方面，从现实影响来看，绿色创新为 ESG 表现奠定坚实基础。环境方面，绿色创新应用帮助企业提高能源和原材料的使用效率，有效减少污染物排放，降低生产过程中的碳足迹；社会方面，绿色创新通过提供健康安全的产品和改善生产条件，加强企业的社会责任感和公众形象；治理层面上，绿色创新要求企业建立一套有效的决策机制和治理结构，公开其环境绩效和创新活动的结果以建立利益相关者信任。总的来说，企业绿色创新能力提升还可以帮助企业履行更多的社会责任，并推动企业管理效率提高、组织结构扁平化，从而提高公司治理能力。企业绿色创新水平的提升可以全面提高企业的 ESG 表现。因此，绿色创新水平的提升会直接影响企业的绿色发展表现，促进企业增长和行业变革。从这一范畴来讲，为促进企业 ESG 发展，应该充分抓住贸易政策不确定性带来的机遇，促进企业绿色创新，推动企业 ESG 方面的综合表现，帮助企业和社会更好实现绿色可持续发展。

表 5.2　机制分析

| | 行业竞争 | 绿色创新 |
|---|---|---|
| | (1) | (2) |
| | Indus_hhi | Patent |
| *L.tpuweigh* | -0.00533*** | 0.461*** |
| | (0.00108) | (0.119) |
| *lnage* | 0.0120*** | -0.0612 |
| | (0.00174) | (0.152) |
| *far* | 0.0132 | -0.707 |
| | (0.00734) | (0.453) |
| *tagr* | -0.000831 | -0.398*** |
| | (0.00222) | (0.0912) |
| *lev* | 0.00150 | 2.513*** |
| | (0.00490) | (0.319) |
| *hindex* | 0.0315*** | 4.225*** |
| | (0.00812) | (0.628) |
| *roa* | -0.0684*** | 7.126*** |
| | (0.0171) | (0.999) |
| _cons | 0.269*** | -5.254*** |
| | (0.0116) | (0.656) |
| *Year FE* | YES | YES |
| *Industry FE* | YES | YES |
| N | 18313 | 14442 |
| $R^2$/Wald | 0.297 | 1464.29*** |

# 第六章　主要结论与政策启示

在当今全球经济格局的多变背景下，尤其是鉴于近年全球新冠疫情对经济和社会的深远影响，跨国交流、企业运营以及民众生活方式均遭遇重大挑战。这种背景下的“不确定性”将成为经济发展的一个持久特征。在这样的不确定性环境中，重构并调整未来的经济发展模式，注重 ESG 的长期可持续性并在此基础上创造长期价值成为下一阶段经济发展的关键。本文基于 2010-2020 年我国 A 股上市企业数据，实证检验贸易政策不确定性对企业 ESG 表现的影响及其作用机制。主要结论如下：

（1）贸易政策不确定性显著提升企业 ESG 表现。在进行一系列稳健性检验和内生性处理后本结论依然成立。由此可见，贸易政策的不稳定性不只触发企业短期应对机制，更是对其长远的战略规划产生根本性影响。表明在贸易政策不确定性环境中，企业强化 ESG 表现是其风险管理和提升市场竞争力的关键策略。

（2）从作用机制来看，贸易政策不确定性通过行业竞争渠道和绿色创新渠道促进企业 ESG 表现提升。其中，行业竞争渠道表现为贸易政策不确定性会加剧企业所处行业的竞争，根据战略性增长期权理论和动态能力理论，在贸易政策不确定性创造的充满高度未知风险的市场中，企业会选择提升 ESG 表现，快速应对环境变化，为自己构筑长期的竞争优势。绿色创新渠道是指贸易政策不确定性增加将激励企业进行绿色技术创新，根据熊彼特的创新理论，企业正是在通过创新来实现自我更新和增长，并使其成为发展的核心动力。

（3）贸易政策不确定性的企业 ESG 表现提升作用对不同类型企业存在差异化影响。第一，从企业业务属性来看，贸易政策不确定性对高新技术企业 ESG 表现的影响更为显著。高新技术企业展现出优异的技术适应性和资源整合能力，通过增加研发和创新投入来巩固国内市场地位和竞争优势，并将 ESG 表现纳入长期战略。第二，从内部控制来看，贸易政策不确定性对企业 ESG 表现的正向影响在内部控制程度较高的企业中显著，但对内部控制程度较低的企业几乎不显著。企业的高质量内部控制有助于降低融资成本，增强财务灵活性，进而支持 ESG 表现提升。第三，从 CEO 环保背景来看，贸易政策不确定性对于 CEO 具有环保背景的企业 ESG 表现提升作用最为显著，这也说明具有环保背景的 CEO 能够引领高管团队关注环保，采纳绿色战略，尤其是在贸易政策不确定性的背景下，对推动企业的绿色发展和提升 ESG 表现具有关键影响。

目前，我国的 ESG 发展处于探索和起步阶段，如何创造完善的外部条件和制度环境，推动 ESG 理念更好地在本土落地实施，仍然是政府极为关心的问题。本文从宏微观层面研究贸易政策不确定性对 ESG 表现的影响及其作用机制。在此基础上，可以提出一系列政策建议，以助政府及相关部门有效应对贸易政策不确定性，并正确监督和引导企业 ESG 行为。本文提出的主要政策建议如下：

（1）完善 ESG 标准体系建设

政府部门在制定 ESG 相关标准时，应根据不同类型、行业企业的特点，针对性地制定 ESG 的披露标准和绩效评价体系。例如，政府可以为高新技术和内部控制健全的企业制定更高的 ESG 标准，而对剩余企业提供各方面支持和指导以助其逐步提升管理水平和 ESG 透明度。

（2）建立 ESG 激励政策

政府可以制定针对性强、效果显著的ESG激励政策，例如，创立专门的ESG基金，用于支持企业的股权和债务融资等活动，并将企业ESG表现纳入其信用评估体系中，并提供相应优惠政策。同时，政府也应对未能达到规定标准的企业作出适当惩罚，以此督促所有企业重视ESG表现。

（3）提高贸易政策透明度和可预测性

政府可以建立一个全面的贸易政策公开平台，定期更新贸易政策的相关动态、解读和法规变更等信息，并提供清晰易懂的解释与指导。通过这样的平台，企业可以及时了解到最新的贸易政策变动贸易政策变动，避免因信息不对称而导致的误解和风险。政府也可以通过与其他国家分享信息和经验，加强协调和合作，以有效降低全球贸易环境的不确定性，为本国企业发展提供更加稳定和可预测的外部环境。

# 参考文献